\documentclass[preprint,12pt,3p,times]{elsarticle}

\usepackage{amssymb}
\usepackage{amsmath}
\usepackage{array}
\usepackage{graphicx}
\usepackage{cases}
\usepackage{multirow}
\usepackage{xcolor}

\usepackage{url,hyperref}
\hypersetup{colorlinks=true,
linkcolor=blue,
citecolor=blue,
filecolor=blue,
urlcolor=blue}

\newcommand*{\Scale}[2][4]{\scalebox{#1}{$#2$}}%
\newcolumntype{P}[1]{>{\centering\arraybackslash}p{#1}}

\journal{Annals of Physics}

\begin{document}

\begin{frontmatter}

\title{Solving the Regge-Wheeler and Teukolsky equations: supervised versus unsupervised physics-informed neural networks}

\author[label1]{Alan S Cornell}
\ead{acornell@uj.ac.za}
\author[label2]{Rhameez S Herbst}
\ead{sherbst@uj.ac.za}
\author[label1]{Anele M Ncube\corref{cor1}}
\ead{ancube@uj.ac.za}
\cortext[cor1]{Corresponding author}
\author[label1]{Hajar Noshad}
\ead{hnoshad@uj.ac.za}
\affiliation[label1]{organization={University of Johannesburg, Department of Physics},
             addressline={PO Box 524, Auckland Park},
             city={Johannesburg},
             postcode={2006},
             state={Gauteng},
             country={South Africa}}

\affiliation[label2]{organization={University of Johannesburg, Department of Mathematics and Applied Mathematics},
             addressline={PO Box 524, Auckland Park},
             city={Johannesburg},
             postcode={2006},
             state={Gauteng},
             country={South Africa}}

\begin{abstract}
To expand on the burgeoning research on physics-informed neural networks (PINNs) and their ability to solve the eigenvalue problems in black hole (BH) perturbation theory, we implement a supervised learning approach to solve the Regge-Wheeler and Teukolsky equations, the equations of gravitational perturbations of Schwarzschild and Kerr BHs, respectively. To date, applications of PINNs using the data-free (unsupervised) learning approach have proven their ability to compute quasinormal mode frequencies of BHs, quantities with physical significance in gravitational wave astronomy. To investigate the potential use of PINNs to compute quasinormal mode overtones higher than the low-lying $n=0$ and $n=1$ modes (with $n$ indexing overtones), the present work has instead applied the supervised approach to simplify computations. Consistent with the universal approximation theory of neural networks, it is found that the PINN algorithm has the intrinsic ability to recover the complex frequencies for various spin sequences (i.e. $s=-2$, $a \in \{0.1, 0.2, 0.3, 0.4\}$, $\ell = 2$, $m \in \{0, 1, 2\}$, $n \in \{0, 1, 2, 3, 4\}$), with approximation errors increasing with the rotation parameter $a$ and overtone number $n$ as a result of the residuals from the training data.
\end{abstract}


\end{frontmatter}

\section{Introduction}
\label{sec:introduction}

\par Since 1963, the Kerr black hole (BH) has been thought to best represent a rotating BH solution~\cite{PhysRevLett.11.237} to Einstein's field equations, a spacetime parameterized by two physical quantities: a mass ($M$) and a non-zero rotation parameter ($a$)~\cite{Teukolsky_2015}. Its relevance to experiment and observation became established more recently by the confirmation of a source of gravitational waves (GWs), the GW150914 signal produced by a pair of Kerr BHs mutually orbiting as a binary system \cite{PhysRevLett.116.061102}. Modeling the GWs produced by such a system before coalescence (i.e., ``merger'') can involve numerically solving Einstein's field equations, which is an arduous task, particularly for non-zero BH rotation and extreme mass ratio cases \cite{Walker_2023}. Fortunately, the stage following the merger, namely the ``ringdown'', is well-described by a linearly perturbed Kerr BH, an effective one-body model involving a Kerr spacetime traversed by a perturbing test particle \cite{PhysRevLett.123.111102}. The ``ringing'' of a perturbed spacetime metric refers to the superposition of damped oscillations known as quasinormal modes (QNMs) that dominate the post-merger part of a GW signal. Determining QNMs and the associated countably infinite, discrete QNM spectrum of characteristic frequencies involves BH perturbation theory and determining the eigenmodes of a non-Hermitian operator \cite{RevModPhys.83.793}.

\par Naturally, QNMs are finding practical use in GW astronomy to estimate the parameters of sources in the high-frequency range of the GW spectrum -- the range of detection of the existing terrestrial GW observatories, namely LIGO, Virgo, KAGRA, and GEO600 \cite{Olaf_Dreyer_2004, PhysRevD.97.044048}. The effectiveness of using QNM frequencies to identify the parameters of BHs stems from their exclusive dependence on BH parameters, irrespective of the various astrophysical processes that can lead to the formation of a perturbed BH (e.g. coalescence events or supernovae) \cite{konoplya2011quasinormal}. Therefore, the inference of a BH's properties encoded in the spectrum of QNM frequencies has been dubbed ``BH spectroscopy'' by several works \cite{Olaf_Dreyer_2004, PhysRevLett.123.111102, PhysRevD.98.084038, PhysRevLett.117.101102}. As shown in Ref. \cite{PhysRevX.9.041060}, BH spectroscopy for an astrophysical binary BH source becomes more effective when several frequencies that match in the dominant spherical harmonic (i.e. $(\ell, m) =  (2,2)$), but differ in overtone number $n$, are superimposed to take advantage of the high signal-to-noise ratio at early times during the post-merger stage. The overtone number $n$ orders the QNM frequencies (denoted by $\omega_n$) according to the damping rate (represented by the imaginary part of $\omega_n$), with the least damped overtone being the $(\ell, m, n) = (2, 2, 0)$ mode. Any method used to compute the QNM spectrum must compute accurate QNM frequencies for several overtones (at least up to $n = 7$) such that high precision estimates of GW source properties can be obtained \cite{PhysRevX.9.041060}. 

\par The relevant differential equation in the case of a perturbed Kerr BH is the Teukolsky equation \cite{1973ApJ...185..635T}. When variable separation is carried out, the Teukolsky equation yields a system of two second-order ordinary differential equations -- one with radial dependence and the other with angular dependence. Both are eigenvalue problems, but the eigenpairs of the radial equation have the most physical relevance to GW astronomy; that is, the eigenfunctions are associated with the Weyl scalar describing outgoing radiation ($\psi_4$), and the eigenvalues are QNM frequencies. The relevance of $\psi_4$ is due to its equivalence to the sum of the time derivatives of the ``plus'' and ``cross'' polarizations of gravitational radiation \cite{10.1063/1.1704788}. On the other hand, the eigenpairs of the angular equation are spheroidal harmonics as eigenfunctions and the separation constants ($A_{\ell m}$) as eigenvalues. Determining $\omega_n$ and $A_{\ell m}$ is the task of several existing numerical and semi-analytic techniques, including Leaver's continued fraction method (CFM) \cite{leaver1985analytic}, the WKB method \cite{PhysRevD.41.374, PhysRevD.68.024018}, the asymptotic iteration method~\cite{10.1155/2012/281705}, spectral and pseudospectral methods \cite{Ripley_2022}. See Ref. \cite{konoplya2011quasinormal} for a comprehensive review of methods in BH perturbation theory. 

\par Recently, physics-informed neural networks (PINNs) have been demonstrated as another viable method from their ability to accurately compute the $n = 0$ spin sequences of the QNM spectrum for asymptotically-flat perturbed Schwarzschild \cite{PhysRevD.106.124047} and Kerr BHs \cite{PhysRevD.107.064025} (both cases involve analytically intractable differential equations, i.e. the Regge-Wheeler and Teukolsky equations). PINNs are good partial differential equation solvers whose abilities as universal function approximators make them extensively applicable. Note that various kinds of partial differential equations can be solved using PINNs, including integro-differential equations, fractional differential equations, forward problems, and inverse problems~\cite{RAISSI2019686, lu2021deepxde}. The ``residuals'' are used as the error (or ``loss'') of an optimization algorithm \cite{RAISSI2019686}. In approximating a solution function to a differential equation, they may apply a primarily data-driven ``supervised learning'' approach, where the values of the functions are not only known at the boundaries of a domain of interest (in the form of Dirichlet, Neumann, Robin or periodic boundary conditions \cite{lu2021deepxde}) but examples of the ``target'' function within the domain are also given. This facilitates the estimation of the mean square error of the network that is minimized in a standard optimization algorithm. An alternative to supervised learning is ``unsupervised learning,'' where there is no data about the solution function within the domain; only the differential equation and boundary conditions are known. This last approach is more useful to solve eigenvalue problems, like the Schr\"{o}dinger equation \cite{jin2020, 9891944, 10.1063/5.0161067}, to which the equations of BH perturbations are similar \cite{konoplya2011quasinormal}. However, given the multiplicity of eigenpairs satisfying both the differential equation and boundary conditions of an eigenvalue problem, there is an as yet generally unresolved challenge of dynamically constraining the optimization to determine, using PINNs, the different QNM frequency overtones as training progresses. 

\par To further motivate this work to construct suitable neural network optimization constraints, especially for determining the overtones in the QNM spectrum of the Teukolsky equation, the current work considers a simpler though still non-trivial task of simultaneously interpolating between the data of the eigenfunction sampled from the relevant domain and approximating the associated eigenvalues. This supervised learning approach, detailed in Sec.~\ref{subsection: unsupervised} is used to test the accuracy of PINN-computed overtones and compare them with the performance of extant numerical methods. As such, the Regge-Wheeler and Teukolsky equations for which these tests are done are outlined in Sec.~\ref{section: theory}. Sec.~\ref{section: physics-informed neural networks} describes the PINN approach, and Sec.~\ref{section: results} demonstrates the setting up of PINNs to compute overtones for several spin sequences. We close with a discussion in Sec.~\ref{section: discussion} on the implications of our results for future applications of PINNs in BH perturbation theory. 

\section{Black hole perturbation theory: asymptotically-flat space-times}
\label{section: theory}
\subsection{Schwarzschild black hole}
\label{section: Schwarzschild}

\par In general, to obtain the equations governing the propagation of fields of various spins in some BH metric, $g_{\mu\nu}$, we consider their respective equations of motion. Furthermore, astrophysical boundary conditions are imposed on the associated solutions (i.e. the QNMs) such that they are ingoing at the BH event horizon and outgoing at spatial infinity. In the temporal domain, they generally decay exponentially. To outline BH perturbation theory, we begin with the Schwarzschild case whose metric is given as~\cite{1916AbhKP1916..189S}:
\begin{equation}
ds^2 = g^{0}_{\mu\nu} dx^{\mu}dx^{\nu} = -f(r)dt^2 + f(r)^{-1}dr^2 + r^2d\theta^2 + r^2\sin^2{\theta}d\phi^2,
\end{equation}

\noindent where $f(r) = 1 - 2M/r$ is the Schwarzschild metric function and $M$ is the BH mass in geometric units. We consider a perturbating quantity $h_{\mu\nu}$, which is small compared to the BH metric (i.e. $h_{\mu\nu} \ll g^{0}_{\mu\nu}$), such that the metric tensor of a linearly perturbed metric is given as~\cite{PhysRev.108.1063}:
\begin{equation}
\label{metric tensor}
    g_{\mu\nu} = g^{0}_{\mu\nu} + h_{\mu\nu}.
\end{equation}
\noindent Here $g^{0}_{\mu\nu}$ denotes (in Schwarzschild coordinates) the unperturbed metric.

\par The equations for gravitational linear perturbations of a Schwarzschild BH arise from Einstein's field equations with $g_{\mu\nu}$ given by Eq.~(\ref{metric tensor}). As shown in Ref.~\cite{1973ApJ...185..635T}, odd and even parity perturbations (axial and polar perturbations, respectively) are obtained from generalizing to tensor quantities the well-known development in spherical harmonics established for scalars, vectors and spinors. Through this development, a system of second-order differential equations is obtained that can be pared down to a single differential equation, given (using the units where: $c = G = 1$) as~\cite{1973ApJ...185..635T}:
\begin{equation}
\label{master equation}
\frac{d^2 Q}{d r_*^2} + \Bigg[\omega^2 - V(r)\Bigg]Q = 0,
\end{equation}

with 
\begin{equation}
\label{effective potential}
 \hspace{-0.5cm}   V(r) = 
    \begin{cases}
    f(r)\left[\dfrac{\ell(\ell + 1)}{r^2} - \dfrac{6M}{r^3}\right]\ \textnormal{(odd parity)};\\ 
    \\
    \dfrac{2f(r)}{r^3} \dfrac{(9M^3 + 9M^2r + 3Mr^2 + 2r^3)}{(3M + r)^2}\ \textnormal{(even parity)}.
    \end{cases}    
\end{equation}

\noindent Here $Q$ denotes the QNM wavefunction, and $r_*$ is a tortoise coordinate related to the radial coordinate $r$, which in this case can be expressed as:
\begin{eqnarray}
\label{tortoise coordinate}
r_*(r) = r + 2M\ln{\left(\frac{r}{2M} - 1\right)}.
\end{eqnarray}

Since the corresponding QNM frequencies of even- and odd-parity wavefunctions are isospectral~\cite{konoplya2011quasinormal}, it suffices to compute only the odd-parity QNMs. The asymptotic behavior of QNMs wavefunctions $Q$ (as dictated by the boundary conditions of an asymptotically-flat metric) is expressed as~\cite{PhysRevLett.52.1361}:
\begin{equation}
\label{boundary conditions}
\lim_{r_* \to -\infty} Q(r_*)= e^{-i\omega r_*};\qquad \lim_{r_* \to \infty} Q(r_*) = e^{i\omega r_*}.
\end{equation}

\subsection{Kerr black hole}
\label{section: Kerr}

\par For gravitational perturbations of a Kerr metric, as was done for the Schwarzschild BH, we consider changes to the metric tensor (Eq.~(\ref{metric tensor})). The derivation was undertaken in Ref.~\cite{1973ApJ...185..635T} using the Newman-Penrose formalism, which enabled a separability of variables in the perturbation equations in spite of the absence of spherical symmetry.

\par The separated, source-free (vacuum) equations for perturbations of a Kerr BH by fields of ``spin weight'' $s$ are obtained by considering the field quantity $\psi$, separated in spherical coordinates according to~\cite{1973ApJ...185..635T}:
\begin{equation}
    \psi = e^{-i\omega t}e^{im\phi}S(\theta)R(r).
\end{equation}

Furthermore, $\psi$ is dependent on a physically measurable quantity, the Weyl scalar $\psi_4$, according to $\psi = \rho^{-4}\psi_{4}$, with $\rho = -1/(r - ia \cos{\theta})$ and $a$ being the rotation parameter. The Weyl scalar is, in turn, given by $\psi_4 = \ddot{h}_{+} + i\ddot{h}_{\times}$; that is, a sum of the second-order time derivatives of the ``plus'' $h_{+}$ and ``cross'' $h_{\times}$ polarizations of gravitational radiation. The separated perturbation equations are~\cite{1973ApJ...185..635T, leaver1985analytic}:
\begin{equation}
\label{radial ODE}
     \Delta^{-s}\frac{d}{dr}\left(\Delta^{s+1}\frac{R}{r}\right) + \left( \frac{K^2 - 2is(r - M)K}{\Delta} + 4is\omega r - \lambda \right) R = 0,\\
\end{equation}
\begin{equation}
\label{angular ODE}
    \frac{d}{du}\left[(1 - u^2)S\right] + \left[a^2\omega^2u^2 - 2a\omega s u + s + A - \frac{(m + su)^2}{1 - u^2} \right] S = 0,
\end{equation}

\noindent with $K = (r^2 + a^2)\omega - am$, $\lambda = A + a^2\omega^2 - 2a m \omega$, $\Delta = r^2 - 2Mr + a^2 = (r - r_{+})(r - r_{-})$, $r_{\pm} = M \pm \sqrt{M^2 - a^2}$ and $u = \cos{\theta}$. Imposing regularity at $\theta = 0$ and $\theta =\pi$ on the angular equation (Eq.~(\ref{angular ODE})) yields an eigenvalue problem, with the separation constants $A = A_{\ell m}$ being eigenvalues, and the spheroidal harmonics $S$ being the eigenfunctions. In the Schwarzschild limit (i.e. $a = 0$) the separation constant reduces to:
\begin{equation}
    \lim_{a \rightarrow 0} A_{\ell m} = \ell (\ell + 1) - s (s + 1).
\end{equation}

\par For the radial equation (Eq.~(\ref{radial ODE})), the relevant boundary conditions (i.e. imposing ingoing sinusoids at the outer horizon of the Kerr BH and outgoing sinusoids at spatial infinity) are expressed as~\cite{PhysRevD.107.064030}:
\begin{equation}
    \lim_{r_*\rightarrow -\infty} R = \frac{e^{ikr_*}}{\Delta^s},\quad \lim_{r_*\rightarrow \infty} R = \frac{i\omega r_*}{r^{2s + 1}},
\end{equation}

\noindent where $k= \omega - m\Omega_{H}$, with $\Omega_{H}$ as the angular frequency of the outer horizon (i.e. $\Omega_{H} = a/(2Mr_{+})$). Here the tortoise coordinate $r_*$ is given by $dr_* = (r^2 + a^2)/\Delta dr$ or:
\begin{equation}
    r_* = r + \frac{r^2_{+} + a^2}{r_{+} - r_{-}}\ln{\left(\frac{r - r_{+}}{r_{+}} \right)} - \frac{r^2_{-} + a^2}{r_{+} - r_{-}}\ln{\left(\frac{r - r_{-}}{r_{+}}\right)}.
\end{equation}

\par In solving the Teukolsky equation using the CFM, Ref.~\cite{leaver1985analytic} utilised an ansatz for $R(r)$ that satisfies the boundary conditions and is given as:
\begin{equation}
    R(r) = e^{i\omega r} (r - r_{-})^{-1-s+i\omega + i\sigma_{+}}(r - r_{+})^{-s-i\sigma_{+}} f(r).
\end{equation}

\noindent Here $\sigma_{+} = (\omega r_{+} - am)/2\sqrt{M^2 - a^2}$. To ensure numerical computations are carried out in a finite domain, we carry out a change in variables from the physical radial domain $r \in [r_+, \infty)$ to a new domain $x \in [0, 1]$ by setting $x = r_+/r$. As such, we have: 
\begin{equation}
\label{radial BCs}
    R(x) = e^{i\omega r_+/x} \left(\frac{r_+}{x} - r_{-}\right)^{-1-s+i\omega + i\sigma_{+}}\left(\frac{r_+}{x} - r_{+}\right)^{-s-i\sigma_{+}} f(x).
\end{equation}

\par For the angular differential equation, the required regularity at $\theta = 0$ and $\theta = \pi$ is embedded in the ansatz:
\begin{equation}
\label{angular BCs}
    S(u) = e^{a\omega u}(1 + u)^{\vert m - s\vert/2}(1 - u)^{\vert m + s\vert/2}  g(u).
\end{equation}

\par With Eqs.~(\ref{radial BCs}) and~(\ref{angular BCs}), the radial and angular differential equations can be expressed in terms of $f(x)$ and $g(u)$, respectively. This ensures the equations implicitly incorporate the boundary conditions. When applying PINNs to solve the Teukolsky equation, the radial and angular equations in terms of $f(x)$ and $g(u)$ are embedded within the loss function by using neural networks to approximate the eigenfunctions $f(x)$ and $g(u)$, whose derivates are computed exactly (up to working precision) with the use of automatic differentiation. We set $2M = 1$ for all our numerical computations.

\section{Physics-informed neural networks}
\label{section: physics-informed neural networks}

\par PINNs utilize the mechanism underlying standard neural networks, which involves the same elements as a regression problem: a ``trial function'' with tunable parameters, an error (i.e. loss) function, and a dataset to fit the model. While the analogs of a trial function and error exist in PINNs, they generally do not require a dataset for constraining the network model~\cite{lu2021deepxde}. More broadly speaking, physical constraints can be imposed directly on the PINNs by incorporating the differential equations in the loss function instead of enforcing them empirically using a training dataset. With the differential equation in the loss function as a neural network regularizer, the approximate solution of the given differential equation (denoted by $\hat{f}(x)$) is generated by the neural network output through the minimization of the loss function (using the same optimization algorithm as is done within standard deep learning)~\cite{RAISSI2019686}. For the specific case of eigenvalue problems, the differential equations to be solved generally take the form:
\begin{equation}
\label{eigenvalue problem}
\mathcal{D}f(x) = \lambda f(x),    
\end{equation}

\noindent where $\mathcal{D}$ is some differential operator, while $f(x)$ and $\lambda$ denote the eigenfunctions and eigenvalues, respectively. The output of PINNs is a composite function constructed by linear transformations acting on layers of parameter vectors, as expressed by the recurrence relation:
\begin{equation}
\label{linear transformation}
\mathcal{N}^{\ell}(\mathbf{x}) =  \sigma\left(\mathcal{N}^{\ell- 1}(\mathbf{x})\mathbf{W}^{\ell} + \mathbf{b}^{\ell}\right),\qquad \textnormal{for}\qquad 1 \leq \ell \leq L.
\end{equation}

The output layer $\mathcal{N}^{\ell=L}(\mathbf{x})$ represents the eigenfunction $f(x)$ in Eq.~(\ref{eigenvalue problem}). In general, for each network layer $\ell$, $\mathcal{N}^{\ell}(\mathbf{x})$ is an output vector produced by multiplying an ``input'' vector (the previous layer's output) $\mathcal{N}^{\ell- 1}(\mathbf{x})$ with a matrix of weights $\mathbf{W}^{\ell}$ and adding a bias matrix $\mathbf{b}^{\ell}$ (see Eq.~(\ref{linear transformation})). The quantities $\mathbf{W}^{\ell}$ and $\mathbf{b}^{\ell}$ signify tunable parameters whose values are arbitrarily initialized before the network is trained. They are subsequently updated in the course of training when the approximated $\hat{f}(x)$ is tuned towards the target solution $f(x)$. Furthermore, $\sigma$ in Eq.~(\ref{linear transformation}) represents a non-linear activation function applied to each hidden layer of the network (excluding the output layer). This facilitates the approximation of a broad family of differentiable functions, as per the universal approximation theory of neural networks~\cite{ lu2021deepxde, 1999AcNum...8..143P}.

\par The PINN algorithm entails a series of forward and backward passes through the network. The randomly initialized weight and bias matrices have elements selected from probability distributions, such as the uniform distribution: $\mathcal{U}(-\sqrt{k}, \sqrt{k})$, where $k = 1/N$ and $N$ is the number of input layer neurons. Each cycle of forward and backward passes propagates input data through the network and backpropagates the gradients of the loss function, which are used to update the network weight and bias matrices in each optimization step. For example, with the stochastic gradient descent (SGD) algorithm, a standard optimization algorithm, each update carries out the following transformations:
\begin{eqnarray}
\label{SGD1}
w^{\ell} &\rightarrow& w^{\ell} - \frac{\eta}{m} \sum_x \frac{\partial \mathcal{L}_x}{\partial w^{\ell}},\\
\label{SGD2}
b^{\ell} &\rightarrow& b^{\ell}  - \frac{\eta}{m} \sum_x \frac{\partial \mathcal{L}_x}{\partial b^{\ell}},
\end{eqnarray}

\noindent where $w^{\ell}$ and $b^{\ell}$ signify each weight and bias element in a given layer $\ell$, and the summations are over all the $m$ input data points. This constitutes a minibatch of a dataset utilized in one given forward-backward pass and subsequent update step. The term $\eta$ is a small, positive parameter, the learning rate, that determines how quickly the weights and biases are updated. A variation of the SGD algorithm, and a common optimization algorithm, is the adaptive moment estimation (Adam) optimizer proposed by Ref.~\cite{kingmaBa2017}.

\par A crucial feature that enables the implementation of the aforementioned algorithms and the relevant differential equation is automatic differentiation. With this method, it is possible to compute the exact numerical values of the derivatives used within backpropagation and the optimization algorithm (Eqs.~(\ref{SGD1}) and (\ref{SGD2})). This capability to numerically compute derivatives is taken advantage of in PINNs when we incorporate differential equations, such as Eq.~(\ref{eigenvalue problem}), into the loss function without the need for discretization (as is carried out with traditional numerical methods).

\subsection{Unsupervised physics-informed neural networks}
\label{subsection: unsupervised} 

\par The idea of using unsupervised (i.e. data-free) PINNs to solve eigenvalue problems was adopted in Ref.~\cite{jin2020}, where they used the approach to solve a quantum Sturm-Liouville problem involving the 1D time-independent Schr\"{o}dinger equation. In our previous work (Ref.~\cite{PhysRevD.106.124047}) we adopted this approach to solve the radial equation governing perturbations of a Schwarzschild BH (whose form resembles the 1D Schr\"{o}dinger equation insofar as being a second-order linear differential equation and an eigenvalue problem). In that case, the differential equation is non-Hermitian and is satisfied by a spectrum of complex-valued eigenvalues.

\par Unsupervised neural networks differ from the standard supervised setup by enforcing physics constraints on the neural network model. In the former, we impose differential equations and associated boundary conditions as constraints, while in the latter, the constraints are in the form of a labeled dataset that implicitly captures the physical laws. In both scenarios, the constraints are subsumed in the loss function to incorporate the physics knowledge in the optimization algorithm. Both constraints can be included in the loss simultaneously or used exclusively.

\par In our implementation of this approach in Ref.~\cite{PhysRevD.106.124047}, the solutions we obtained had the same level of numerical precision as the standard approaches, such as the CFM~\cite{leaver1985analytic} that was used as a reference. However, there remained the challenge of learning overtones (indexed by $n$). This challenge is also encountered by the recent work by Luna {\it et al.}~\cite{PhysRevD.107.064025}, who generated the QNMs of Kerr BHs for a small part of the QNM spectrum (i.e. only the fundamental mode and first overtone ($n = 1$)), where the use of ``seed values'' was found to be not sufficiently constraining to drive the learning of QNMs with $n > 1$. 

\par Constructing sufficiently strong constraints to enable accurate searches for overtones using an unsupervised approach is a challenging task that requires modifications of the loss functions. A simple first step is to determine the ability of PINNs to learn QNM overtones from data used as additional physics constraints, which leads us to test the supervised approach.

\subsection{Supervised physics-informed neural networks}
\label{subsection: supervised} 

\par When a supervised approach is utilized, PINNs solve inverse-type problems that are boundary-value or initial-value problems with unknown parameters (e.g. $\lambda$ in Eq.~(\ref{eigenvalue problem}))~\cite{lu2021deepxde}. Note, though, that we know the solution function partially through a sampling of training data at some points in the domain. This is the opposite of a problem of the forward type, where the differential operator and boundary conditions are well-defined, but the solution is entirely unknown. In this case, the loss function is given, in general, as~\cite{lu2021deepxde}:
\begin{equation}
\mathcal{L}(\mathbf{\theta}; \mathcal{T}) =  w_{DE}\mathcal{L}_{DE}(\mathbf{\theta}; \mathcal{T}) + w_{BC}\mathcal{L}_{BC}(\mathbf{\theta}; \mathcal{T}) + w_{data}\mathcal{L}_{data}(\mathbf{\theta}; \mathcal{T}),  
\end{equation}

\noindent where $w_{DE}, w_{BC}, w_{data}$ are user-defined weights while $\mathbf{\theta} = \{\textbf{W}^{\ell}, \textbf{b}^{\ell}\}_{1 \leq \ell \leq L}$ represents the weights and biases collectively. Training points randomly selected from the computational domain are denoted by $\mathcal{T}$ and the functions $\mathcal{L}_{DE}$ and $\mathcal{L}_{BC}$ are the differential equation residual and boundary condition losses, respectively:
\begin{eqnarray}
\label{PDEloss}
\mathcal{L}_{DE}(\mathbf{\theta}; \mathcal{T}) &=&  \Scale[1]{\frac{1}{|\mathcal{T}_{DE}|} \sum_{\mathbf{x} \in \mathcal{T}_{DE}} \Vert f(\mathbf{x}; \frac{\partial \hat{u}}{\partial x_1}, \ldots, \frac{\partial \hat{u}}{\partial x_d} ; \frac{\partial^2 \hat{u}}{\partial x_1 \partial x_1}, \ldots, \frac{\partial^2 \hat{u}}{\partial x_d \partial x_d} ; \hat{\mathbf{\lambda}} ) \Vert^2_2,}\\
\mathcal{L}_{BC}(\mathbf{\theta}; \mathcal{T}) &=& \frac{1}{|\mathcal{T}_{BC}|} \sum_{\mathbf{x} \in \mathcal{T}_{BC}} \Vert \mathcal{B}(\hat{u}, \mathbf{x})\Vert^2_2.
\end{eqnarray}

Here, the ``hat'' notation of $\hat{u}$ and $\hat{\mathbf{\lambda}}$ signify the network's approximations of the variable $u(\mathbf{x})$ and any unknown coefficients or eigenvalues within the differential equations. The loss term $\mathcal{L}_{data}$ represents, for inverse problems, the mean square error of $\hat{u}(\mathbf{x})$ compared to a ``labeled dataset'' of the response variable $u(\mathbf{x})$:
\begin{equation}
\label{observational}
\mathcal{L}_{data}(\mathbf{\theta}; \mathcal{T}) = \frac{1}{|\mathcal{T}_{data}|} \sum_{\mathbf{x} \in \mathcal{T}_{data}} \Vert u(\mathbf{x}) - \hat{u}(\mathbf{x}) \Vert^2_2.    
\end{equation}

\par As such, inverse problems require a dataset of target values $u(\mathbf{x})$ for optimization. Applying this approach to the Regge-Wheeler and the Teukolsky equations involves the generation of a dataset of the QNM eigenfunctions using other numerical techniques. For our work, we utilized the CFM~\cite{leaver1985analytic} implemented in Mathematica ~\cite{gritgravitationintecnico} to generate datasets consisting of $\sim 10^{3}$ points from the 1D intervals making up the domains of the perturbation equations. 

\par During optimization, the eigenvalues of the differential equation are inferred from the associated eigenfunctions. These are represented by the networks in the typical way that inverse problems are solved~\cite{RAISSI2019686}. The performance of this approach is limited in part by the approximation errors inherited from the CFM that was used for data generation. Nevertheless, our results (see Sec.~\ref{section: results}) indicate that the universal approximation theory of neural networks is valid when approximating the QNMs of BHs. That is, given sufficiently well-defined physical constraints, neural network approximations can recover many different, physically allowed eigenstates that make up the eigenspectrum of a given eigenvalue problem.

\par In solving the perturbation equations of interest, we utilize the same notation used by Leaver~\cite{leaver1985analytic}. Regarding the gravitational perturbations of a Schwarzschild BH, we consider solving:
\begin{equation}
\label{Regge-Wheeler equation}
    F_{0} f^{\prime\prime}(\xi) + F_{1} f^{\prime}(\xi) + F_{2} f(\xi) = 0,
\end{equation}

with 
\begin{eqnarray}
    F_{2} &=& -(\ell(1 + \ell) + \epsilon (\xi - 1) - 4 \rho(\xi - 2 \rho + \xi\rho - 1 )),\\
    F_{1} &=& (1 + 2\rho - 4\xi(1 + 2 \rho) + \xi^2 (3 + 4 \rho)),\\
    F_{0} &=& (\xi - 1)^{2} \xi.
\end{eqnarray}

Here, we have carried out a change in coordinates from $r \in (1, \infty)$ to $\xi \in (0, 1)$ using the transformation $\xi = (r - 1)/r$. This transformation is necessary for obtaining a finite domain that facilitates solving the problem numerically. The spin-weight of the perturbing field $s$ is incorporated in Eq.~(\ref{Regge-Wheeler equation}) through the term $\epsilon = s^2 - 1$, where $\epsilon = 3$ for gravitational fields. Furthermore, the QNM frequency is denoted by $\rho = -i\omega$ and the boundary conditions for QNMs of a Schwarzschild BH (Eq.~(\ref{boundary conditions})) are automatically satisfied by Eq.~(\ref{Regge-Wheeler equation}), which was obtained by imposing the ansatz:
\begin{equation}
    \psi = e^{-\xi\rho/\xi - 1} 
    (1 / 1 - \xi)^{2\rho} (\xi / 1 - \xi)^{\rho}  f(\xi).
\end{equation}

\par Evaluating Eq.~(\ref{Regge-Wheeler equation}) at given points in the computational domain specified by the input data points, and taking the mean of the square, gives $\mathcal{L}_{DE}$ in this case. $\mathcal{L}_{data}$ is then the mean square error of the PINN approximations $\hat{f}(\xi)$ compared with the associated target values provided by the dataset of points $f(\xi)$, within the domain of $\xi$.

\begin{figure}
\begin{center}
\includegraphics[scale=0.8]{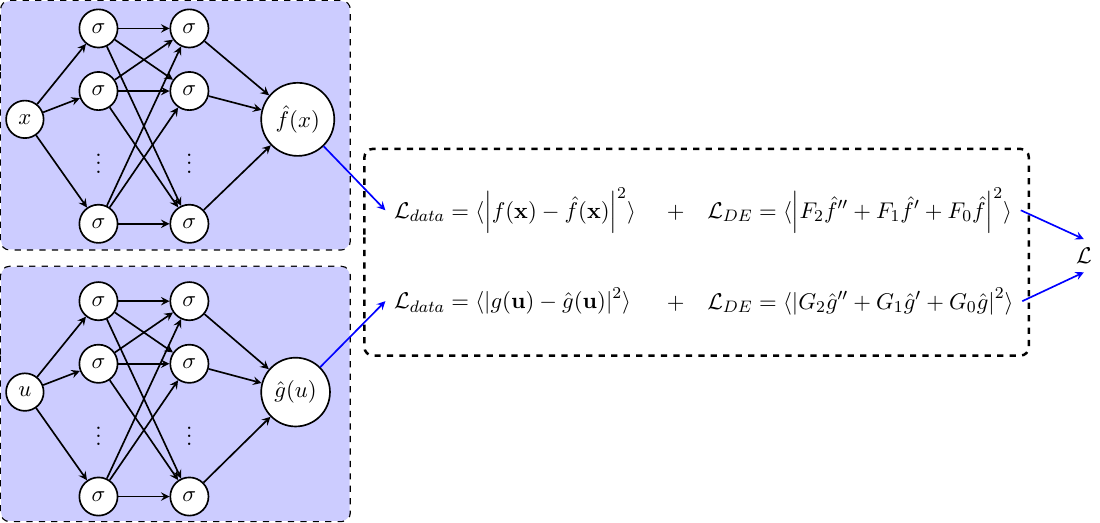}
\caption{\label{figure 1}  A schematic of the PINN used to approximate the functions $f(x)$ and $g(x)$, and their derivatives, such that the governing Eqs.~(\ref{modified Teukolsky equation}) and labeled data are satisfied.}
\end{center}
\end{figure}

\par For the Teukolsky equation, where the boundary conditions are implicitly incorporated in Eqs.~(\ref{radial BCs}) and (\ref{angular BCs}), we use the finite domains $x \in [0, 1]$ and $u \in [-1, 1]$. As was done in Ref.~\cite{PhysRevD.107.064025}, Eqs.~(\ref{radial ODE}) and (\ref{angular ODE}) can be given in terms of $f(x)$ and $g(u)$ in the form:
\begin{equation}
\label{modified Teukolsky equation}
    F_2 f^{\prime\prime} + F_1f^{\prime} + F_0f = 0,\qquad G_2 g^{\prime\prime} + G_1g^{\prime} + G_0g = 0,
\end{equation}

\noindent where the coefficients $F_i = F_{0, 1, 2}$ and $G_i = G_{0,1,2}$ are functions of $x$ and $u$, respectively (see Ref.~\cite{PhysRevD.107.064025} for the full expressions). Solving the Teukolsky equation with PINNs is facilitated by constructing two separate neural networks to approximate $f(x)$ and $g(u)$ and, in turn, satisfy Eqs.~(\ref{modified Teukolsky equation}). The setup is illustrated in Fig.~\ref{figure 1}, where the mean squares of the differential equations are incorporated into the loss function, in addition to the mean square errors that take into account the labeled data. A similar but simpler setup applies when using the supervised PINNs to compute the QNMs of a Schwarzschild BH.

\par Regarding the implementation of PINNs, we use the PyTorch library extensively to code the tensors, linear transformation, activation functions, and automatic differentiation required to implement learning algorithms. The code was run on Google Colaboratory's Intel(R) Xeon(R) on 12GB of RAM, which was sufficient for the computations needed in this work.

\section{Eigenvalues of the differential equations governing gravitational perturbations of\\ Schwarzschild and Kerr black holes}
\label{section: results}

\begin{figure}[ht!]
\includegraphics[width=0.9\linewidth]{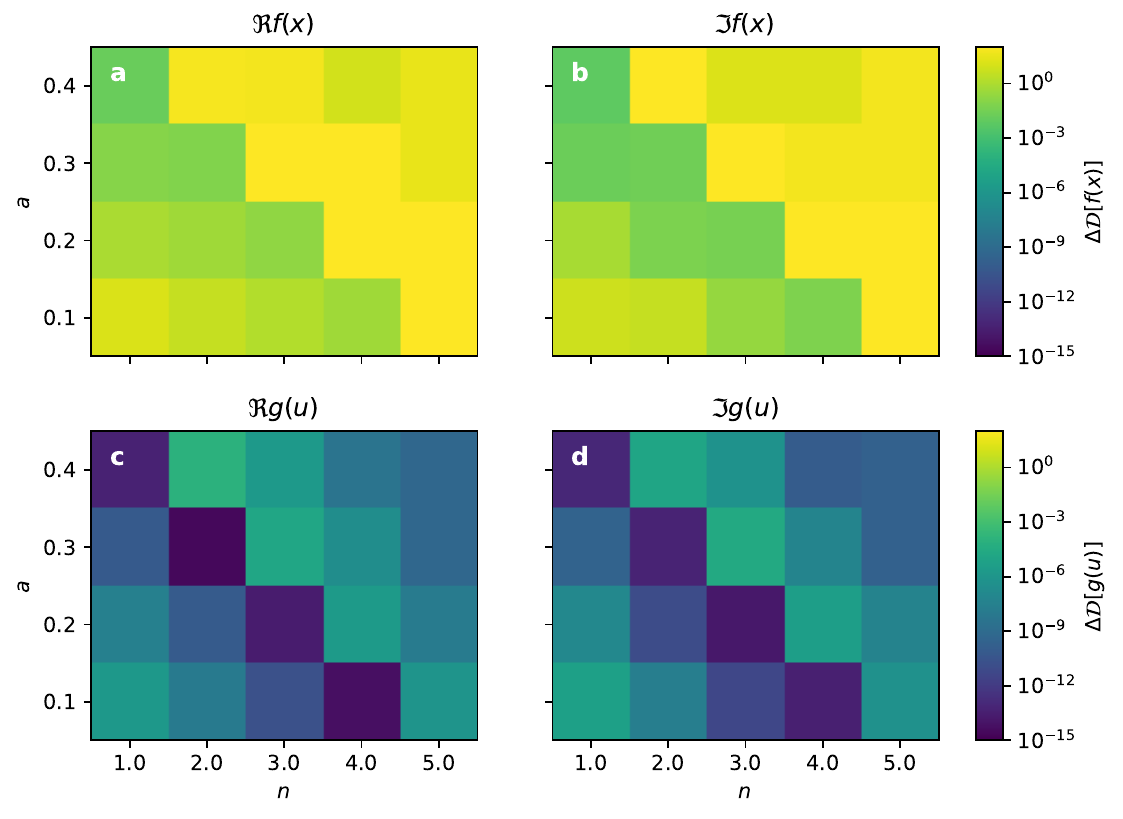}
\caption{\label{residuals} With $\mathcal{D}$ generically representing a differential operator, plots of the average residuals $\Delta\mathcal{D}(f(x))$ and $\Delta\mathcal{D}(g(x))$ due to the approximation errors in the CFM computations of the complex eigenfunctions $f(x)$ and $g(u)$. These are presented against the overtone number $n$ and the rotation parameter $a$ (see Ref.~\cite{gritgravitationintecnico} for the implementation of the CFM in Mathematica).}
\end{figure}

\par In this section, the QNM eigenvalues are computed using supervised PINNs. These results show that, given a grid of labelled data points selected from the computational domain (i.e. the radial and angular domains relevant to the Teukolsky equation), PINNs can simultaneously determine the eigenfunctions that concur with input data and the associated QNM frequencies not known beforehand. Note that the precomputed dataset that facilitated supervised optimization of the PINNs consisted of complex-valued eigenfunctions (i.e. vectors $f(x = x_i)$ and $g(u = u_i)$ at discrete training points $x_i$ and $u_i$, respectively) whose accuracy diminishes with increasing QNM overtone number ($n$) and rotation parameter ($a$). This is an artifact of the CFM, where the diminishing accuracy is reflected by the residuals of the radial- and angular-dependent differential equations (see Fig.~\ref{residuals} where the residuals are denoted by $\Delta \mathcal{D}(f(x))$ and $\Delta \mathcal{D}(g(u))$). These are associated with the CFM approximations of $f(x)$ and $g(u)$ for various spin sequences (i.e. QNMs specified by indices $a$, $s$, $\ell$, $m$, $n$). The residuals (which ideally should be zero) are the averages of the absolute values of the left-hand-sides of the radial and angular differential equations (i.e. Eq.~(\ref{modified Teukolsky equation})) evaluated at points sampled from the intervals $x \in [0, 1]$ and $u \in [-1, 1]$. They give a proxy of the inaccuracy of the CFM-computed eigenfunctions whose values at the sampled data points are $f(x_i)$ and $g(u_i)$ in the expressions:
\begin{equation}
\Delta \mathcal{D}(f(x)) = \frac{1}{|\mathbf{x}|}\sum^{|\mathbf{x}|}_{i=1}|\mathcal{D}(f(x_i))|,\quad \Delta \mathcal{D}(g(u)) = \frac{1}{|\mathbf{u}|}\sum^{|\mathbf{u}|}_{i=1}|\mathcal{D}(g(u_i))|.
\end{equation}

\noindent Here $\mathbf{x} = \{x_1, ..., x_{|\mathbf{x}|}\}$ and $\mathbf{u} = \{u_1, ..., u_{|\mathbf{u}|}\}$, where $|\mathbf{x}|$ and $|\mathbf{u}|$ are total number of points sampled from the radial and angular domains. The CFM-approximated eigenfunctions are, in turn, determined from the ansatzes~\cite{leaver1985analytic}:
\begin{equation}
f(x) \approx \sum^{N}_{n = 0} d_n \left[\frac{(r_{+}/x) - r_{+}}{(r_{+}/x) - r_{-}}\right]^n,\quad g(u) \approx \sum^{N}_{n = 0} a_n \left(1 + u\right)^n,
\end{equation}

\noindent where $N$ is some arbitrarily large real value; $a_n$ and $d_n$ are expansion coefficients (``minimal solution sequences'') that are determined during the implementation of the CFM. As seen in the following results, the dependence of the accuracy of the CFM-computed eigenfunctions with $a$ and $n$ carries over to the performance of the PINNs that use the CFM approximations as training data. As such, we have constrained the space of spin sequences to low-lying but non-zero $a$ and $n$, which we considered to be a good starting point for demonstrating the ability of neural networks to approximate overtones of BH with non-zero spin.

\par The PINN computed QNMs are compared with the CFM computed values given in Ref.~\cite{leaver1985analytic}, where the same set of spin sequences (and some additional ones) for which the QNM frequencies were determined (using the latter method) is used to generate the PINN approximated QNM frequencies. These are the QNMs of Schwarzschild BHs with indices: $a = 0, \ell = 2, m = 0, n \in \{0, 1, 2, 3, 4\}$; the fundamental mode QNMs frequencies of non-zero-spin Kerr BHs: $a \in \{0.0, 0.1, 0.2, 0.3, 0.4\},\ell = 2, m \in \{0, 1\}, n = 0$; and the overtones of non-zero-spin Kerr BHs: $a \in \{0.0, 0.1, 0.2, 0.3, 0.4\}, \ell = m = 2$ (the dominant angular mode, as noted in Ref.~\cite{PhysRevX.9.041060}), $n \in \{1, 2, 3, 4\}$ (see Tables~\ref{Kerr overtones a = 0.0},~\ref{Kerr QNMs a > 0},~\ref{Kerr overtones a = 0.1} and~\ref{Kerr overtones a = 0.3}, respectively). Note that we use the same units as in Ref.~\cite{leaver1985analytic}, $c = G = 2M = 1$. As such, the extremal limit of the rotation parameter is $a = 0.5$.

\par Regarding the size of the precomputed datasets, for each set of indices in the space of spin sequences specified above, we generated a grid of $\sim$ 900 points in the radial domain $x \in [0, 1]$ and $\sim$ 2000 points in the angular domain $u \in [-1, 1]$; that is, $\mathbf{x} =\{x_1, ..., x_{900}\}$, $\mathbf{u} \in \{u_1, ..., u_{2000}\}$ for the associated complex-valued eigenfunctions $f(\mathbf{x})$ and $g(\mathbf{u})$. However, in the actual training the datasets are shuffled, balanced (yielding 900 points in both radial and angular domains), and split into training, validation, and testing sets using a ratio of 8:1:1. Note that as outlined in Ref.~\cite{2021NatRP...3..422K}, the learning algorithm utilized in supervised PINNs combines two types of physics constraints on the neural networks, namely ``observation biases'' and ``learning biases''. The former entails interpolation between points in labeled training data, while the latter are the differential equations that need to be satisfied at points in the domain where they are evaluated. In our case, the Teukolsky equation constitutes the learning biases without which the neural networks cannot determine the QNM frequencies and separation constants (not present in the training data) even with the labeled data. This distinguishes PINNs from an exclusively interpolating model.
\begin{center}
\begin{table}[ht!]
\caption{\label{Kerr overtones a = 0.0} Spin sequences: $a=0$ (``Schwarzschild limit''), $\ell = 2$, $m = 0$, $n \in \{0, 1, 2, 3, 4\}$.\newline}
\begin{tabular}{P{0.05\linewidth}P{0.05\linewidth}P{0.4\linewidth}P{0.4\linewidth}}
\hline
                       &                       & CFM~\cite{Stein:2019mop}            & PINN \\
$\ell$                 & $n$                   & $\omega_{\textsubscript{$n$}}$      & $\hat{\omega}_{\textsubscript{$n$}}$ \\
                       &                       &                                     & ($\Delta\Re\hat{\omega}_{\textsubscript{$n$}}$, $\Delta\Im\hat{\omega}_{\textsubscript{$n$}}$)\normalsize\\
\hline\\[0.05em]
\multirow{2}{0.5em}{2} & \multirow{2}{2em}{0.0}& \multirow{2}{10em}{0.747343 - 0.177925i} & 0.747806 - 0.177663i\\
                       &                                          &                       & (0.062\%, -0.147\%) \\[0.5em]
                       & \multirow{2}{2em}{1.0}& \multirow{2}{10em}{0.693422 - 0.547830i} & 0.693669 - 0.547440i\\
                       &                                          &                       & (0.036\%, -0.071\%) \\[0.5em]
                       & \multirow{2}{2em}{2.0}& \multirow{2}{10em}{0.602107 - 0.956554i} & 0.602604 - 0.956577i\\
                       &                                          &                       & (0.083\%, 0.002\%)  \\[0.5em] 
                       & \multirow{2}{2em}{3.0}& \multirow{2}{10em}{0.503010 - 1.410296i} & 0.486670 - 1.390154i\\
                       &                                          &                       & (-3.248\%, -1.428\%)\\[0.5em]
                       & \multirow{2}{2em}{4.0}& \multirow{2}{10em}{0.415029 - 1.893690i} & 0.392655 - 1.832018i\\
                       &                                          &                       & (-5.391\%, -3.257\%)\\[0.5em]
\hline
\end{tabular}
\end{table}
\end{center}

\begin{table}[ht!]
\caption{\label{Kerr QNMs a > 0} Spin sequences: $a \in \{0.1, 0.2, 0.3, 0.4\}$, $\ell = 2$, $m \in \{0, 1\}$, $n = 0$.\newline}
\begin{tabular}{P{0.05\linewidth}P{0.05\linewidth}P{0.225\linewidth}P{0.275\linewidth}P{0.275\linewidth}}
\hline
                       &                        & CFM~\cite{Stein:2019mop}            & \multicolumn{2}{c}{PINN} \\ 
 $m$                   & $a$                    & $\omega_{\textsubscript{$0$}}$      & $\hat{\omega}_{\textsubscript{$0$}}$   & $\hat{A}_{\textsubscript{$2 m$}}$ \\[0.05em]
                       &                        & \{$A_{\textsubscript{$2 m$}}$\}     & ($\Delta\Re\hat{\omega}_{\textsubscript{$0$}}$, $\Delta\Im\hat{\omega}_{\textsubscript{$0$}}$) & ($\Delta\Re \hat{A}_{\textsubscript{$2 m$}}$, $\Delta\Im \hat{A}_{\textsubscript{$2 m$}}$) \\
\\
\hline\\[0.05em]
\multirow{2}{2em}{0}   & \multirow{2}{2em}{0.1} & 0.7502 - 0.1774i      & 0.7507 - 0.1765i       & 3.9973 + 0.0013i     \\
                       &                        & \{3.9972 + 0.0014i\}  & (0.055\%, -0.500\%)    & (0.001\%, -0.006\%)  \\[0.05em]
                       & \multirow{2}{2em}{0.2} & 0.7594 - 0.1757i      & 0.7623 - 0.1720i       & 3.9885 + 0.0055i     \\
                       &                        & \{3.9886 + 0.0056i\}  & (0.390\%, -2.098\%)    & (-0.001\%, -0.008\%) \\[0.05em]
                       & \multirow{2}{2em}{0.3} & 0.7761 - 0.1720i      & 0.7825 - 0.1633i       & 3.9727 + 0.0121i     \\
                       &                        & \{3.9730 + 0.0126i\}  & (0.821\%, -5.067\%)    & (-0.007\%, -0.053\%) \\[0.05em]
                       & \multirow{2}{2em}{0.4} & 0.8038 - 0.1643i      & 0.8088 - 0.1441i       & 3.9464 + 0.0185i     \\
                       &                        & \{3.9480 + 0.0223i\}  & (0.619\%, -12.315\%)   & (-0.039\%, -0.365\%) \\[0.05em]
\hline\\[0em]
\multirow{2}{2em}{1}   & \multirow{2}{2em}{0.1} & 0.7765 - 0.1770i      & 0.7769 - 0.1757i       & 3.8930 + 0.0250i     \\
                       &                        & \{3.8932 + 0.0252i\}  & (0.057\%, -0.722\%)    & (-0.003\%, -0.021\%) \\[0.05em]
                       & \multirow{2}{2em}{0.2} & 0.8160 - 0.1745i      & 0.8161 - 0.1722i       & 3.7673 + 0.0519i     \\
                       &                        & \{3.7676 + 0.0532i\}  & (0.015\%, -1.305\%)    & (-0.007\%, -0.123\%) \\[0.05em]
                       & \multirow{2}{2em}{0.3} & 0.8719 - 0.1691i      & 0.8788 - 0.1598i       & 3.6086 + 0.0788i     \\
                       &                        & \{3.6125 + 0.0835i\}  & (0.785\%, -5.526\%)    & (-0.107\%, -0.432\%) \\[0.05em]
                       & \multirow{2}{2em}{0.4} & 0.9605 - 0.1559i      & 0.9749 - 0.1280i       & 3.3906 + 0.0926i     \\
                       &                        & \{3.4023 + 0.1122i\}  & (1.504\%, -17.910\%)   & (-0.342\%, -1.762\%)\\[0.05em]
\hline
\end{tabular}
\end{table}

\begin{table}[ht!]
\caption{\label{Kerr overtones a = 0.1} Spin sequences: $a \in \{0.1, 0.2\}$, $\ell = m = 2$, $n \in \{ 1, 2, 3, 4\}$.\newline}
\begin{tabular}{P{0.05\linewidth}P{0.05\linewidth}P{0.225\linewidth}P{0.275\linewidth}P{0.275\linewidth}}
\hline
                       &                        & CFM~\cite{Stein:2019mop}              & \multicolumn{2}{c}{PINN} \\ 
 $a$                   & $n$                    & $\omega_{\textsubscript{$n$}}$        & $\hat{\omega}_{\textsubscript{$n$}}$    & $\hat{A}_{\textsubscript{$22$}}$ \\[0.05em]
                       &                        & \{$\hat{A}_{\textsubscript{$2 2$}}$\} & ($\Delta\Re\hat{\omega}_{\textsubscript{$n$}}$, $\Delta\Im\hat{\omega}_{\textsubscript{$n$}}$) & ($\Delta\Re \hat{A}_{\textsubscript{$22$}}$, $\Delta\Im \hat{A}_{\textsubscript{$22$}}$) \\
\\
\hline\\[0.05em]
\multirow{2}{2em}{0.1} & \multirow{2}{2em}{1.0} & 0.7580 - 0.5411i      & 0.7616 - 0.5432i      & 3.7924 + 0.1548i       \\
                       &                        & \{3.7958 + 0.1504i\}  & (0.484\%, 0.397\%)    & (-0.091\%, 0.386\%)    \\[0.5em]
                       & \multirow{2}{2em}{2.0} & 0.6786 - 0.9357i      & 0.6878 - 0.9447i      & 3.8201 + 0.2614i       \\
                       &                        & \{3.8222 + 0.2589i\}  & (1.357\%, 0.954\%)    & (-0.054\%, 0.196\%)    \\[0.5em]
                       & \multirow{2}{2em}{3.0} & 0.5888 - 1.3653i      & 0.5968 - 1.3859i      & 3.8413 + 0.3648i       \\
                       &                        & \{3.8543 + 0.3758i\}  & (1.363\%, 1.511\%)    & (-0.339\%, -0.801\%)   \\[0.5em]
                       & \multirow{2}{2em}{4.0} & 0.5050 - 1.8177i      & 0.4386 - 2.0858i      & 0.692137 + 0.063640i   \\
                       &                        & \{3.8880 + 0.4980i\}  & (-13.148\%, 14.752\%) & (-82.198\%, -28.994\%) \\[0.5em]
\hline\\[0em]
\multirow{2}{2em}{0.2} & \multirow{2}{2em}{1.0} & 0.8417 - 0.5295i      & 0.8622 - 0.5395i      & 3.5274 + 0.3156i       \\
                       &                        & \{3.5386 + 0.3091i\}  & (2.437\%, 1.900\%)    & (-0.315\%, 0.502\%)    \\[0.5em]
                       & \multirow{2}{2em}{2.0} & 0.7756 - 0.9065i      & 0.809542 - 0.949656i  & 3.5731 + 0.5436i       \\
                       &                        & \{3.5935 + 0.5253i\}  & (4.382\%, 4.759\%)    & (-0.568\%, 1.202\%)    \\[0.5em]
                       & \multirow{2}{2em}{3.0} & 0.6977 - 1.3076i      & 0.7324 - 1.3834i      & 3.6277 + 0.6778i       \\
                       &                        & 3.6654 + 0.7509i      & (4.971\%, 5.799\%)    & (-1.028\%, -4.177\%)   \\[0.5em]
                       & \multirow{2}{2em}{4.0} & 0.6231 - 1.7235i      &  0.6139 - 2.0312i     & 1.0488 + 0.0235i       \\
                       &                        & \{3.7462 + 0.9810i\}  & (-1.473\%, 17.852\%)  & (-72.003\%, -48.334\%) \\[0.5em]
\hline
\end{tabular}
\end{table}

\begin{table}[ht!]
\caption{\label{Kerr overtones a = 0.3} Spin sequences: $a \in \{0.3, 0.4 \}$, $\ell = m = 2$, $n \in \{1, 2, 3, 4\}$.\newline}
\begin{tabular}{P{0.05\linewidth}P{0.05\linewidth}P{0.225\linewidth}P{0.275\linewidth}P{0.275\linewidth}}
\hline
                       &                        & CFM~\cite{Stein:2019mop}            & \multicolumn{2}{c}{PINN} \\ 
 $a$                   & $n$                    & $\omega_{\textsubscript{$n$}}$      & $\hat{\omega}_{\textsubscript{$n$}}$    & $\hat{A}_{\textsubscript{$22$}}$ \\[0.05em]
                       &                        & \{$A_{\textsubscript{$22$}}$\}      & ($\Delta\Re\hat{\omega}_{\textsubscript{$n$}}$, $\Delta\Im\hat{\omega}_{\textsubscript{$n$}}$) & ($\Delta\Re \hat{A}_{\textsubscript{$22$}}$, $\Delta\Im \hat{A}_{\textsubscript{$22$}}$) \\
\\
\hline\\[0.05em]
\multirow{2}{2em}{0.3} & \multirow{2}{2em}{1.0} & 0.9596 - 0.5077i     & 1.0143 - 0.5321i       & 3.1351 + 0.5042i       \\
                       &                        & \{3.1882 + 0.4726i\} & (5.697\%, 4.803\%)     & (-1.665\%, 2.142\%)    \\[0.5em]
                       & \multirow{2}{2em}{2.0} & 0.9084 - 0.8606i     & 0.9979 - 0.9671i       & 3.1694 + 0.8988i       \\
                       &                        & \{3.2697 + 0.7944i\} & (9.860\%, 2.368\%)     & (-3.067\%, 5.819\%)    \\[0.5em]
                       & \multirow{2}{2em}{3.0} & 0.8440 - 1.2263i     & 0.9454 - 1.4408i       & 3.3773 + 0.8002i       \\
                       &                        & \{3.3822 + 1.1194i\} & (12.019\%, 7.494\%)    & (-0.146\%, -15.061\%)  \\[0.5em]
                       & \multirow{2}{2em}{4.0} & 0.7798 - 1.5950i     & 0.8907 - 1.8264i       & 3.8702 + 0.2024i       \\
                       &                        & \{3.5130 + 1.4392i\} & (14.221\%, 14.507\%)   & (10.169\%, -50.706\%)  \\[0.5em]
\hline\\[0em]
\multirow{2}{2em}{0.4} & \multirow{2}{2em}{1.0} & 1.1558 - 0.4563i     & 1.2702 - 0.5238i       & 2.4699 - 0.7320i       \\
                       &                        & \{2.6313 + 0.6171i\} & (9.891\%, 14.782\%)    & (-6.135\%, 7.106\%)    \\[0.5em]
                       & \multirow{2}{2em}{2.0} & 1.1245 - 0.7678i     & 1.3976 - 1.5137i       & 3.7667 - 0.0703i       \\
                       &                        & \{2.7230 + 1.0312i\} & (24.292\%, 97.152\%)   & (38.331\%, -47.307\%)  \\[0.5em]
                       & \multirow{2}{2em}{3.0} & 1.0779 - 1.0858i     & 1.4128 - 1.5089i       & 3.9130 - 0.2256i       \\
                       &                        & \{2.8613 + 1.4431i\} & (31.069\%, 38.970\%)   & (36.756\%, -49.837\%)  \\[0.5em]
                       & \multirow{2}{2em}{4.0} & 1.0125 - 1.3959i     & 1.4413 - 1.7847i       & 5.8168 - 0.0004i       \\
                       &                        & \{3.0464 + 1.8282i\} & (42.346\%, 27.849\%)   & (90.942\%, -64.630\%)  \\[0.5em]
\hline
\end{tabular}
\end{table}

\begin{figure}
\includegraphics[width=0.85\linewidth]{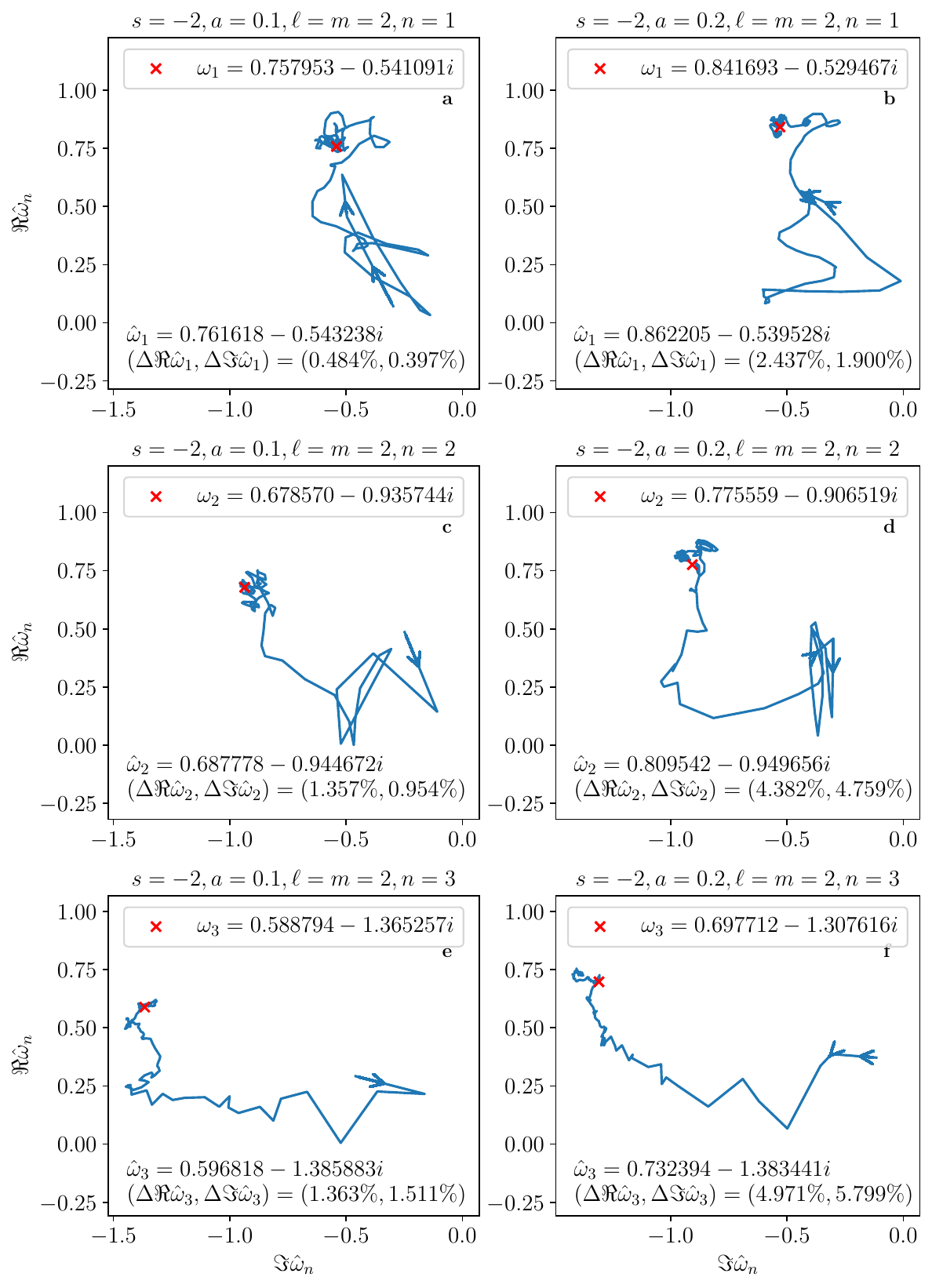}
\caption{\label{overtones} The evolution of the PINN approximated QNM frequencies ($\hat{\omega}_n$), where the arrows point in the direction of the increasing number of training epochs and the blue line traces the values of $\hat{\omega}_n$ from some arbitrary starting point (due to random weight initialization). The line continues up to some optimized value of $\hat{\omega}_n$ close to the target (i.e. $\omega_n$ denoting the CFM values) at the end of training.}
\end{figure}

\begin{figure}
\includegraphics[width=0.9\linewidth]{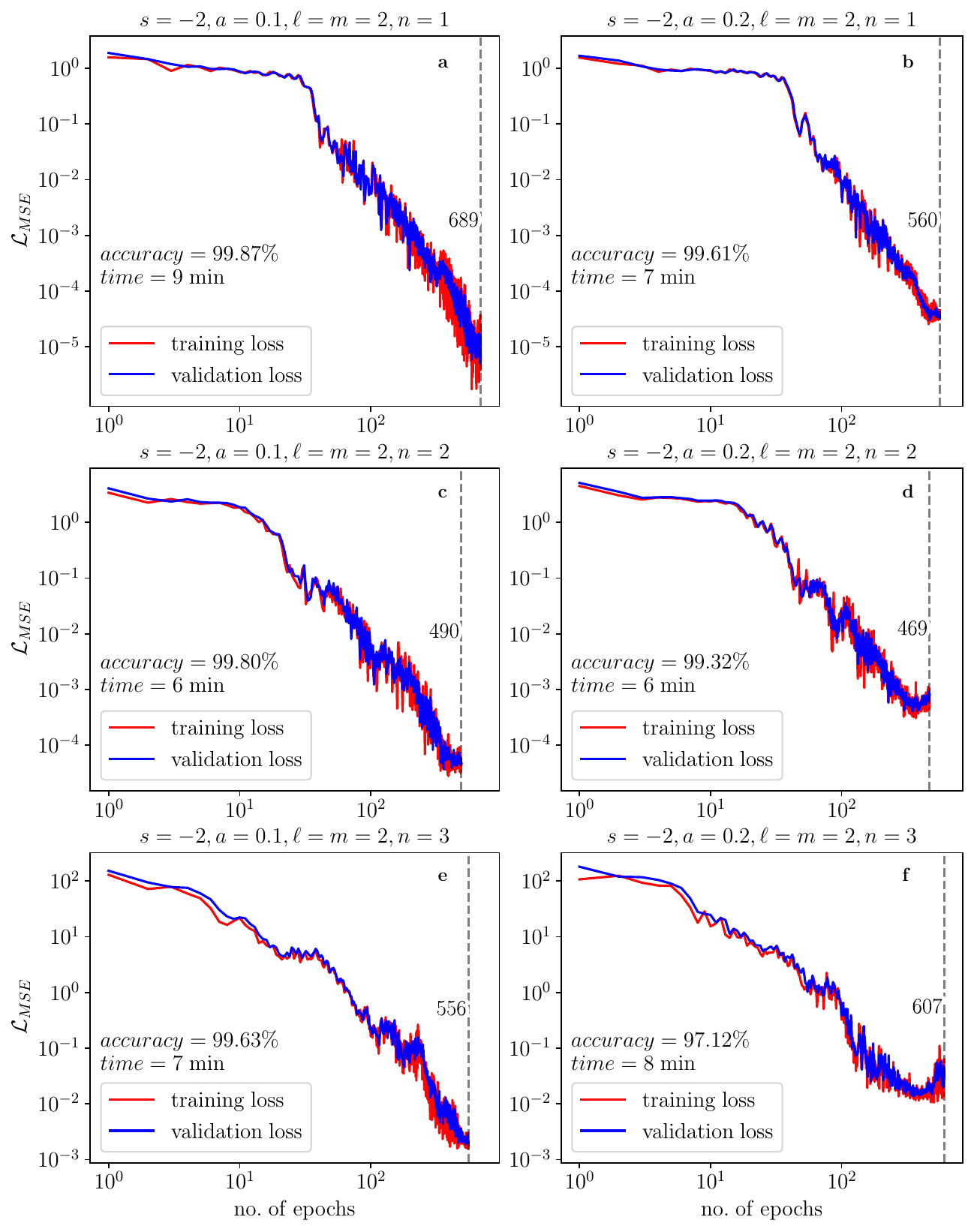}
\caption{\label{losses} Mean square error loss ($\mathcal{L}_{MSE}$) histories for some example spin sequences. The test accuracy of the optimal $\hat{f}(x)$ and $\hat{g}(u)$ is denoted by ``accuracy''; training time before early stopping is denoted by ``time''; and the epoch at which early stopping occurs is indicated by number labeling the dashed line.}
\end{figure}

\par The PINNs used for solving the Teukolsky equation constitute two fully connected, feed-forward neural networks for approximating the two solution eigenfunctions $f(x)$ and $g(u)$. Regarding the specific setups for each of the networks, 3 and 2 hidden layers were used for the $f(x)$ and $g(u)$ networks, respectively, where 20 nodes per layer were used in both cases (the two networks ``talk'' to each other via standard learnable parameters: $\omega_{n}$ and $A_{\ell m}$, the QNM frequencies and separation constants). The two networks utilize a self-scalable tanh activation function~\cite{2022arXiv220412589G}, chosen because of its tunability during training. It is given as:
\begin{equation}
    \sigma^{k}_{\ell} = \tanh{(\mathbf{x})} + \beta^{k}_{\ell} \mathbf{x} \tanh{(\mathbf{x})},
\end{equation}

\noindent where $k$ and $\ell$ indicate the neuron and layer numbers, respectively. The tunable parameter $\beta^{k}_{\ell}$ is akin to the weights and bias parameters within Eq.~(\ref{linear transformation}) and is likewise updated during the optimization.

\par The network architectural setup stated above was applied to solving Eq.~(\ref{modified Teukolsky equation}) in both the Schwarzschild limit (which amounts to computing the QNMs of the Regge-Wheeler equation) and the non-zero-spin cases (i.e. $a > 0$). Importantly, solving the Teukolsky equation in the numerically amenable form (i.e. Eq.~(\ref{modified Teukolsky equation})) with $a=0$ is numerically a different and more challenging task to solve than solving the Regge-Wheeler equation, even though the former reduces to the latter for $a=0$, and the QNMs are equivalent. This indicates the importance (within PINNs) of choosing between different differential equation formulations of a given physical problem.  

\par Concerning the optimization hyperparameters, we tested different setups of the learning rate (with or without learning rate scheduling) and the use of the Adam optimizer, solely or paired with the limited-memory Broyden–Fletcher–Goldfarb–Shanno optimizer (L-BFGS). Outlined here is one optimal setup we found, though hyperparameter optimization was not carried out systematically (which may be explored in the future). For our current purposes, we found it sufficient to use a standard setup consisting of the Adam optimizer with a constant learning rate of 0.01, without a learning rate scheduler such as ReduceLRonPlateau (since it was found to have a detrimental effect on convergence). On the other hand, early stopping was used to end training whenever the validation loss (a proxy of the PINN's overall performance) stopped improving for a span exceeding 100 training epochs, which optimized the training time. This ensured that training ended as soon as convergence was attained and prevented the algorithm training for the maximum $5 \times 10^3$ epochs with $\sim 5$ times the duration. The training time was also optimized by feeding input data in subsets of the full training dataset (i.e. minibatches) of size\footnote{i.e. number of minibatch points} 32 each.

\par Given in Tab.~\ref{Kerr overtones a = 0.0} are the spin sequences of the Teukolsky equations, Eq.~(\ref{modified Teukolsky equation}), in the \newline Schwarzschild limit (as stated previously, this is not an equivalent problem to the Regge-Wheeler equation, Eq.~(\ref{Regge-Wheeler equation})). We compare the values obtained using the PINN approach (i.e. supervised and unsupervised PINNs) with those from the CFM~\cite{leaver1985analytic}. Within our PINN code, we could conveniently utilize the \texttt{qnm} Python open-source library \cite{Stein:2019mop} to provide us with a cache of low-lying spin sequences (generated using Leaver's approach) to use as reference values to determine the accuracy of the PINN computations. The percentage errors in all tables' parentheses reflect the deviations from these references. Our previous work in Ref.~\cite{PhysRevD.106.124047} used unsupervised PINNs to compute only the fundamental mode QNM frequencies. In this work, we find that with the supervised PINNs, it is possible to go further and compute QNM overtones and obtain a broader range of accurately determined values. However, as previously noted (and illustrated in Fig.~\ref{residuals}), the errors in the parentheses increase with $n$. Nonetheless, the ability of PINNs to identify the overtones is striking, albeit only up to about the $n = 3$ limit set by the quality of the dataset used here for demonstration purposes.

\par In Tab.~\ref{Kerr QNMs a > 0}, the fundamental mode frequencies and separation constants of non-zero-spin Kerr BHs are presented. As indicated, the results of Tabs.~\ref{Kerr overtones a = 0.0} and~\ref{Kerr QNMs a > 0} emulate the space of spin sequences generated in Leaver's work~\cite{leaver1985analytic}. Beyond this, additional spin sequences that are astrophysically significant are investigated (see Tabs.~\ref{Kerr overtones a = 0.1} and~\ref{Kerr overtones a = 0.3} for the list of overtones and separation constants of non-zero-spin Kerr BHs, with the dominant angular mode being $\ell = m = 2$~\cite{PhysRevX.9.041060}). Fig.~\ref{overtones} shows corresponding plots of the evolution of the PINN approximations in the complex plane of $\omega_n$. The arrows indicate the direction of increased learning during training before termination due to early stopping (intervals between two consecutive arrows represent a space of 100 epochs). We also include the associated loss histories in Fig.~\ref{losses}, showing the evolution of the mean-square error losses $\mathcal{L}_{MSE}$ (Eq.~(\ref{observational})) during training and validation. The ``test accuracy'' values indicated in the plots are the root-mean-square relative errors of the PINN approximations (i.e. predictions of the trained models on the unseen test set). We can infer from the test accuracies of $\sim 99\%$ that the models are highly accurate interpolators and fit well with the data. The plots for other spin sequences not given here can be found in our GitHub repository~\cite{supervisedKerrPINNs}. Note that the training run times indicated in the plots are specific to the processor used (the indicated Intel Xeon in Google Colab). As such, the training times noted are not a general reflection of the speed with which PINNs can be trained.

\section{Discussion and Conclusions}
\label{section: discussion}

\par This work has presented one way to set up PINNs to solve the eigenvalue problems that arise in BH perturbation theory describing linear perturbations of a Kerr BH with the BH rotation parameter in the range $0 \leq a \leq 0.5$. There is direct applicability of BH perturbation theory in GW astronomy because the direct metric perturbations of the BH's spacetime (i.e. spin $s=-2$ fields) are associated with the observable gravitational radiation produced by BH perturbation events, such as the binary BH coalescences that the LIGO-VIRGO-KAGRA collaborations have detected within the last decade~\cite{PhysRevX.9.031040, PhysRevX.11.021053, PhysRevX.13.041039}. Mathematically, the eigenvalue problems are analogous to the unidimensional time-independent Schr\"{o}dinger equation, but with crucial differences in the astrophysical boundary conditions imposed within the perturbed BH scenario. In many cases, which include the gravitational perturbations of asymptotically-flat Schwarzschild and Kerr BHs, the associated differential equations have no known exact, closed-form solutions, justifying the continual development of approximation methods~\cite{leaver1985analytic, PhysRevD.68.024018, 10.1155/2012/281705}.

\par The PINN algorithm represents one such case of a developing numerical method that is proving useful in solving QNMs of BHs. Notably, in Refs.~\cite{PhysRevD.106.124047, PhysRevD.107.064025}, PINNs were shown to compute the fundamental mode frequencies of asymptotically-flat Schwarzschild and Kerr BHs with values deviating by less than $1\%$ from the reference values within the literature. In principle, neural networks are capable of learning overtones ($n > 0$) as implied by the universal approximation theorem~\cite{HORNIK1989359}. However, a spectrum of QNMs satisfying the same differential equation and boundary conditions presents a difficult challenge for optimization algorithms, such as gradient descent, to systematically search for an individual QNM of interest (among a series of QNMs). Therefore, this work aimed to motivate the creation of neural network architecture and optimization designs to alleviate this impasse. We investigated a simpler problem in which only the QNM frequencies are kept as unknowns while the QNM wavefunctions were known for some points in the finite $x-u$ plane (i.e. $[0,1]\times[-1,1]$). This way, QNM overtones could be learned using the supervised PINN approach, albeit using QNM eigenfunction data with errors incurred from their numerical computation, as described by Fig.~\ref{residuals}.

\par Our results show that supervised PINNs can approximate QNM overtones for the Schwarzchild limit ($a = 0$) and the Kerr case ($a > 0$). However, the imperfections in the training data become prominent for higher overtone numbers and rotation parameter values (see Fig.~\ref{residuals}). This is evident in the corresponding reduction in performance of PINN approximated $\omega_n$ with increasing overtone and spin indices. As expected for a data-driven (supervised) approach to optimization, the PINN approximated solutions are as good as the data used for training. For this reason, the results presented exclude $n > 4$ overtones and $a > 0.4$ spins (with $a = 0.5$ being the extremal limit of the rotation parameter in the $M = 0.5$ units) due to either mounting residuals in the generated data or QNMs not being generated at all for some spin sequences by the reference CFM. Other than the training data, the choice of hyperparameters influenced the accuracy of the PINN approximations and the speed with which convergence occurred. Note, though, that hyperparameters (e.g. number of layers and nodes and learning) were kept fixed for varying $a$ and $n$ in the presented results; therefore, we could attribute the increasing approximation error solely to the CFM-generated data and rule out the contribution by sub-optimal hyperparameter choices. In view of this, we propose that neural networks can potentially compute QNMs (both eigenvalues and eigenfunctions) with as much accuracy in the higher $n$ and $a$ part of the QNM spectrum as in the lower. This could be achieved with a more precisely generated input dataset, where the currently used CFM is used here as it is a standard implementation available to the community.

\par The ability of PINNs to straight-forwardly compute QNMs of a minimally modified (i.e. changing only the coordinate system) form of the Teukolsky equation is its advantage over several existing techniques. Many extant methods are restricted to computing only the eigenvalues and require relatively more complex modifications of the BH perturbation equations. By contrast, computing QNM eigenfunctions with PINNs would be advantageous considering their utility in determining QNM excitation factors related to the detectability of QNMs, see Ref.~\cite{PhysRevD.74.104020}). Considering this and the increasing focus and interest in machine learning applications, there is a lot of incentive to develop PINNs further. Possibilities include creating stand-alone PINN (i.e. unsupervised, data-free PINNs) solvers of BH perturbation equations to compute overtones, with a systematic scanning mechanism built into the loss function (similar to Refs~\cite{9891944, 10.1063/5.0161067}), or merging PINNs with other existing methods, to exploit advantages or complementary strengths of these different methods.

\section*{Acknowledgements}
\noindent ASC is supported in part by the National Research Foundation (NRF) of South Africa; RSH is supported by the CoE-MaSS; AMN is supported by an SA-CERN Excellence Bursary through iThemba LABS. We are very grateful to Anna Chrysostomou for the discussions that went into improving this manuscript.    

\bibliographystyle{elsarticle-num} 
\bibliography{references}

\end{document}